\documentclass{epl}

\title{Optical Zener-Bloch oscillations in binary waveguide arrays}
\author{S. Longhi}

\institute{Dipartimento di Fisica and Istituto di Fotonica e
Nanotecnologie del CNR \\ Politecnico di Milano, Piazza L. da
Vinci, 32 I-20133 Milano (Italy) }


\begin{document}

\maketitle

\begin{abstract}
Zener tunneling in a binary array of coupled optical waveguides
with transverse index gradient is shown to produce a sequence of
regular or irregular beam splitting and beam recombination events
superimposed to Bloch oscillations. These periodic or aperiodic
Zener-Bloch oscillations provide a clear and visualizable
signature in an optical system of coherent multiband dynamics
encountered in solid-state or matter-wave systems.
\end{abstract}

Light propagation in waveguide arrays has provided in the past few
years a conceptually and experimentally relevant model to study
the optical analogue
\cite{Peschel98,Morandotti99,Pertsch99,Lenz99,Longhi05,Trompeter06a,Trompeter06b,Assanto06}
of Bloch oscillations (BO) and related effects typically
encountered in solid-state or matter-wave systems
\cite{Gluck02,Morsch06}. One of the major advantages offered by
waveguide arrays as compared to other optical systems \cite{vari}
is the possibility to easily perform a direct visualization of the
wave packet dynamics in coordinate space \cite{Trompeter06a} which
is complementary to most usual frequency-domain (spectral)
measurements \cite{Gluck02,vari}. Recent experiments in waveguide
arrays \cite{Trompeter06a,Assanto06} have reported the beautiful
observation of optical Zener tunneling (ZT), thus motivating the
study of multiband dynamics in these systems. In presence of a
constant field, which yields a BO motion within a single band
approximation, ZT usually manifests as a cascading of transitions
to higher-order bands which results in the damping of the
oscillatory BO motion or -in the spectral domain- in the
broadening of the Wannier-Stark (WS) resonances
\cite{Gluck02,Callaway74}. In Ref.\cite{Trompeter06a} a direct
visualization of decaying BO via ZT to higher-order bands has been
reported in a polymer waveguide array with a low refractive index
contrast, which is analogous to BO decay observed in Bose-Einstein
condensates in shallow lattices \cite{Morsch06}. In the experiment
of Ref.\cite{Trompeter06a}, the BO motion of a light wave packet
in the first band of the array is periodically damped owing to the
appearance of ZT which manifests itself as regular outbursts of
radiation escaping at the turning points of the motion where the
wave packet reaches the edge of the Brillouin zone in the
reciprocal space. This decay of BO via periodic ZT bursts is
irreversible and involves a cascading of transitions to
higher-order bands. However, as discussed in the context of
semiconductor superlattices \cite{Rotvig95,Hone96}, a coherent
dynamics instead of an irreversible decay process is expected when
only two bands, coupled by ZT, are involved. In this case the
dynamics of occupation probabilities in the two bands shows a
complex coherent behavior which is related to the spacing of the
two interleaved WS ladders \cite{Hone96} and may even lead, under
special conditions, to
suppression of ZT.\\
In this Letter BO dynamics in a binary array of coupled optical
waveguides is theoretically analyzed and shown to provide an ideal
optical system for a direct observation of two-band coherent
Zener-Bloch dynamics, in which ZT manifests itself as an aperiodic
or periodic splitting and recombination of propagating light beam
superimposed to the BO motion. The binary array
\cite{Sukhorukov02} consists of a sequence of alternating
single-mode optical waveguides of different widths $2w_1$ and
$2w_2$, separated by $a/2$, showing the same refractive index peak
change $\Delta n$ [Fig.1(a)]. We describe light propagation in the
array by a rather standard two-dimensional scalar
Schr\"{o}dinger-like equation \cite{Morandotti04,Sukhorukov03},
extended to include a transverse index gradient which simulates
either transverse thermal heating of waveguides
\cite{Trompeter06a} or waveguide curvature \cite{Lenz99,Longhi05}.
By writing the electric field as $E(x,z,t)=\psi(x,z) \exp(ikn_ez-i
\omega t)+c.c.$, where $\omega$ is the angular frequency of the
field, $k=\omega / c_0$ its wave number in vacuum, and $n_e$ an
effective index of the structure \cite{Sukhorukov03}, the
evolution of the field envelope $\psi$ along the propagation $z$
direction satisfies the paraxial wave equation
 \begin{equation}
i \hbar \frac{\partial \psi}{\partial z} = -\frac{\hbar^2}{2n_{e}}
\frac{\partial^2 \psi}{\partial x^2} + [n_e-n(x)] \psi- F x \psi,
 \end{equation}
\begin{figure}
\onefigure[width=10cm]{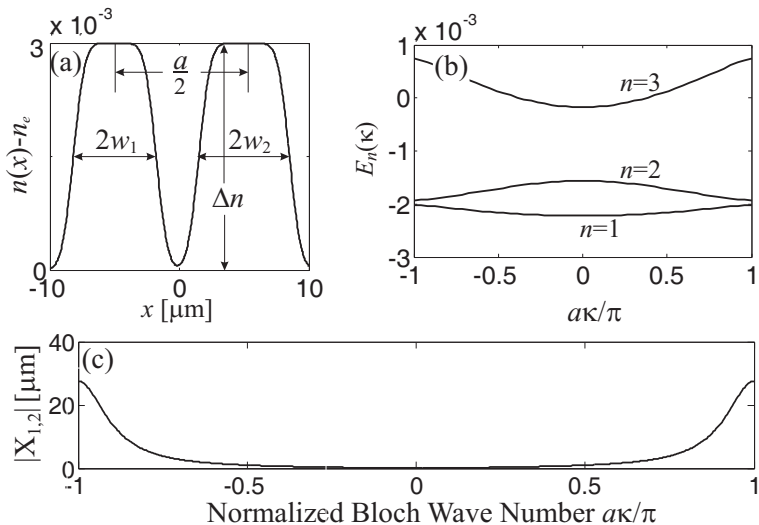} \caption{(a) Effective refractive
index modulation (unit cell) of the binary array used in the
numerical simulations. Parameter values are: $a=20 \; \mu$m,
$w_1=3.2 \; \mu$m, $w_2=3.5 \; \mu$m, $\Delta n=0.003$,
$\lambda=1.44 \; \mu$m, $n_e \simeq 2.138$ (note that $w_1 \neq
w_2$). (b) Band diagram for $F=0$. $n$ is the band index. The
other parameter values are the same as in Fig.1(a). (c) Behavior
of interband coupling coefficient $|X_{1,2}|$}
\end{figure}
\noindent
 where $n(x)-n_e$ is
the effective modulation of the optical refractive index
\cite{Sukhorukov03} in the transverse $x$ direction [with
$|n(x)-n_e| \ll n_e$], $ \hbar= \lambda/ (2 \pi)=1/k$ is the
reduced wavelength, and $F$ is the index gradient coefficient. For
circularly-curved waveguides, $F$ is related to the bending radius
of curvature $R$ by the simple relation $F=n_e/R$ \cite{note0}.
Since the unit cell of the array is a dimer and supports two modes
(namely the symmetric and anti-symmetric combinations of the modes
guided by the narrow and wide waveguides), the band structure
shows two lowest-order bands \cite{Sukhorukov02}, separated by a
gap $E_g$ which vanishes for $w_1 = w_2$. If the gap is large
enough, i.e. if $w_1$ is sufficiently different from $w_2$ and the
index change $\Delta n$ is not too small, LZ is usually negligible
at low or moderate values of $F$ and BO occurs with a spatial
period $z_B= 2 \pi \hbar/(Fa)=\lambda /(Fa)$ (see, for instance,
\cite{Lenz99,Trompeter06a,Lenz03}). However, for $w_1 \simeq w_2$,
the two bands are separated by a small gap and ZT is no more
negligible, although it may be still disregarded for higher-order
bands. A typical band diagram of the binary array \cite{note1} for
$F=0$ and $w_1 \simeq w_2$ is depicted in Fig.1(b) for parameter
values typical of Lithium-Niobate waveguide arrays
\cite{Longhiun}. To study the role of ZT on BO dynamics, we
numerically simulated propagation of a broad Gaussian beam for
different values of the refractive index gradient $F$, i.e. of the
Bloch period $z_B$. For the sake of simplicity, normal incidence
has been assumed, so that at the input plane of the array the
first band ($n=1$) is mostly excited \cite{Morandotti04}. For very
small values of $F$, a characteristic BO motion with period $z_B$
is observed, however as $F$ is increased, i.e. as $z_B$ decreases,
a sequence of beam splitting and beam recombination at planes
$z=z_B/2, 3 z_B/2, 5 z_B/2,...$ is observed, which is related to
ZT as discussed below. The sequence is usually irregular, i.e.
aperiodic, however at some special values of $F$ it shows a
nearly-periodic pattern. Since this ZT sequence is superimposed to
a BO motion with period $z_B$, the resulting beam motion can be
referred to as optical Zener-Bloch oscillations (ZBO).
\begin{figure}
\onefigure[width=10cm]{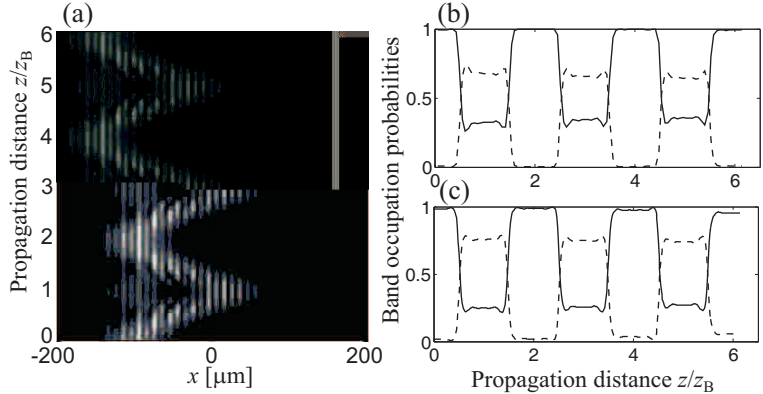} \caption{(a) Gray-scale plot
showing beam intensity propagation in a 8-cm long binary array for
$F=5.52 \; {\rm m}^{-1}$, corresponding to a Bloch period
$z_B=13.04$ mm. The input field is a Gaussian beam
$\psi(x,0)=\exp[-(x+x_0)^2/w_0^{2}]$ with $w_0=45 \; \mu$m and
$x_0=100 \; \mu$m. The geometrical parameters of the binary array
are the same as those of Fig.1. (b) Numerically-computed evolution
of band occupation probabilities $P_1$ and $P_2$ versus
propagation distance. (c) Behavior of $P_1$ and $P_2$ as predicted
by the two-level equations (3) and (4).}
\end{figure}
\begin{figure}
\onefigure [width=10cm]{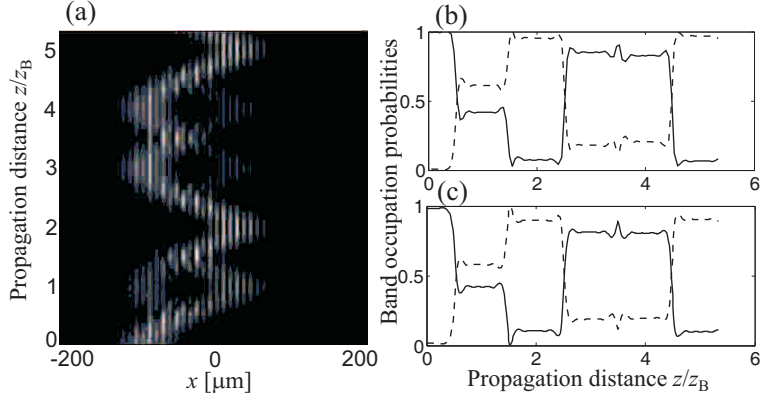}  \caption{Same as Fig.2, but for
$F=4.8 \; {\rm m}^{-1}$ ($z_B=15$ mm), corresponding to aperiodic
ZBO.}
\end{figure}
\noindent As an example, Figs.2(a) and 3(a) show typical numerical
results demonstrating either nearly-periodic or aperiodic ZBO. In
both cases the array length is $8$ cm. Note that, for a curved
waveguide array the values of index gradient $F$ used in the
simulations of Figs.2(a) and 3(a) correspond to a radius of
curvature $R \simeq 38.7$ cm and $R \simeq 44.5$ cm, respectively.
To relate the observed dynamics with ZT, we adopt a rather general
technique developed in solid-state physics \cite{Callaway74} and
expand the wavepacket $\psi(x,z)$ as a superposition of Bloch
states $\varphi_n(x,\kappa)=u_n(x,\kappa) \exp(i \kappa x)$ of the
array, i.e. we set $\psi(x,z)=\sum_n \int_{-\pi /a}^{\pi /a} d
\kappa c_n(z,\kappa) \varphi_n(x, \kappa)$, where $n$ is the band
index, $-\pi/a < \kappa < \pi/a$, $u_n(x+a,\kappa)=u_n(x,\kappa)$,
and the normalization condition $\int_{-\infty}^{\infty} dx
\varphi_{n'}^*(x,\kappa ') \varphi_n(x,\kappa) \equiv \langle
\varphi_{n'}(x,\kappa ') | \varphi_n (x, \kappa) \rangle=
\delta_{n,n'} \delta( \kappa- \kappa')$ holds. The occupation
probability of band $n$ is given by $P_n(z)=\int {d \kappa}
|c_n(z,\kappa)|^2$. For the binary array with the band diagram
shown in Fig.1(b) and for near normal beam incidence, the main
dynamics comprises only the two bands $n=1,2$. Figures 2(b) and
3(b) show the numerically-computed occupation probabilities for
the two lowest-order bands versus propagation distance,
corresponding to the spatial ZBO motion of Figs.2(a) and 3(a).
Note that ZT is characterized by sharp and consistent interband
transitions at planes $z=z_B/2, 3 z_B/2, 5 z_B/2,...$ connected by
plateau regions where ZT is small. In these connecting regions,
the motion of the two wavepackets constructed by the superposition
of Bloch modes belonging to the two lowest-order bands are
independent each other and shows BO. Beam splitting and
recombination is thus due to abrupt power transfer between the two
wavepackets, which in the plateau regions move into opposite
directions according to the semiclassical motion of a Bloch
particle. To get deeper insights into the ZBO dynamics and to
explain the existence of special values of $F$ leading to a
periodic motion, we consider the evolution equations for the
spectral coefficients $c_n(z,\kappa)$, which read
\cite{Callaway74}:
\begin{equation}
i \hbar \frac{\partial c_n}{\partial z}=E_n(\kappa)
c_n(z,\kappa)-iF \frac{\partial c_n}{\partial \kappa}-F \sum_l
X_{n,l} c_l(z, \kappa),
\end{equation}
where $X_{n,l}(\kappa)=X_{l,n}(\kappa)^*=(2 \pi i /a) \int_{0}^{a}
dx \; u^{*}_n(\partial u_l / \partial \kappa)$ and $n,l=1,2$ in
the two-band approximation. The coupling term $X_{1,2}$ in Eq.(2)
accounts for interband transitions, i.e. ZT. For a symmetric index
profile [$n(x)=n(-x)$], the Bloch functions $u_n(x,\kappa)$ can be
taken such that $X_{1,1}=X_{2,2}=0$ and $X_{1,2}(\kappa)=i
\Theta(\kappa)$, with $\Theta$ real valued and
$\Theta(-\kappa)=\Theta(\kappa)$. The numerically-computed
behavior of $X_{1,2}$ for the binary array of Fig.1(a) is shown in
Fig.1(c). For broad beam excitation at normal incidence, the
values of coefficients $c_{1,2}$ at the input plane $z=0$  are
approximately given by $c_1(0,\kappa) \simeq g( \kappa)$ and
$c_2(0,\kappa) \simeq 0$, where the spectrum $g(\kappa)$ is narrow
around $\kappa=0$ . In this case, according to the acceleration
theorem a solution to Eqs.(2) is simply given by
$c_{1,2}(z,\kappa)=f_{1,2}( F z / \hbar) g(\kappa-Fz / \hbar)$,
where $f_{1,2}(\kappa)$ satisfy the two-level equations with
periodic coefficients
\begin{eqnarray}
i F \frac{df_{1}}{d \kappa} & = &  E_{1}(\kappa) f_{1} -i F
\Theta(\kappa) f_{2} \\
i F \frac{df_{2}}{d \kappa} & = &  E_{2}(\kappa) f_{2} +i F
\Theta(\kappa) f_{1}
\end{eqnarray}
with $f_1(0) \simeq 1$ and $f_2(0) \simeq 0$. The occupation
probabilities of the two bands are then given by $P_1(z)=|f_1|^2$
and $P_2(z)=|f_2|^2$. ZT corresponds to non-adiabatic transitions
at the points of avoided crossing between the two bands, i.e at
$\kappa= \pi/a, 3 \pi/a, 5 \pi/a,...$, where the interband
coupling $\Theta$ is largest and the energy separation
$E_2(\kappa)-E_1(\kappa)$ smallest. This explains the
characteristic dynamics shown in Figs.2(b) and 3(b), which is
rather well reproduced by a numerical analysis of the two-level
equations [see Figs.2(c) and 3(c)]. In order to explain the
existence of periodic ZBO at special values of index gradient $F$,
let us note that, according to Floquet theory the solution to
Eqs.(3) and (4) is generally not periodic but characterized by the
two periods $ 2 \pi/a$ and $2 \pi/ |\mu_2-\mu_1|$, where $ i
\mu_{1,2}(F)$ are the Floquet exponents of the system. However,
for special values of $F$ such that $|\mu_2(F)-\mu_1(F)|/a$ is a
fractional number $N/M$ (with $ N$ and $M$ irreducible integers),
the functions $P_{1,2}(z)$ are periodic with period $M z_B$. For
instance, for $|\mu_2-\mu_1|=a/2$ the evolution of occupation
probabilities is periodic with spatial period twice the Bloch
period $z_B$.
\begin{figure}
\onefigure [width=10cm]{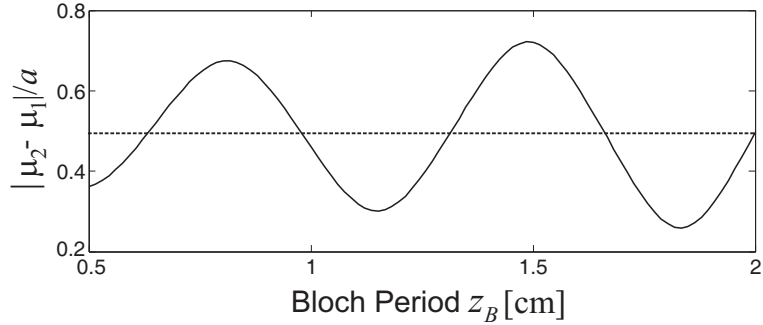} \caption{ Behavior of normalized
Floquet exponent difference $|\mu_2-\mu_1|/a$ versus Bloch period
$z_B$. The dotted curve corresponds to periodic ZBO with period
$2z_B$.}
\end{figure}
\noindent Figure 4 shows the numerically-computed behavior of
$|\mu_2-\mu_1|/a$ versus $z_B= \lambda/(Fa)$. Note that, at the
intersections with the horizontal dashed line, ZBO are periodic
with period $2 z_B$, which is the case shown in Fig.2. The
existence of periodic ZBO is closely related to the spectrum of
Eq.(1), which for a two-band model is given by two interleaved
Wannier ladders whose separation depends on the refractive index
gradient $F$ \cite{Hone96}. We can express the Wannier ladders for
Eq.(1) in terms of the Floquet exponents $i \mu_{1,2}$ of the
two-level equations (3) and (4) by extending the procedure of
Ref.\cite{Callaway74}. We look for solutions of Eqs.(2) in the
form $c_{n}(z,\kappa)= \bar{c}_n(\kappa) \exp(-iEz/ \hbar)$ and
impose the  periodic boundary conditions $\bar{c}_n(0)=\bar{c}_n(2
\pi /a)$ to find
  the  spectrum $E$. After some straightforward calculations, this yields
  for the spectrum the two interleaved Wannier ladders
  $E_{n}=-\mu_1(F)F+nFa$ and $E_{m}=-\mu_2(F)F+mFa$, where $m$ and $n$ are
  arbitrary integers and $i \mu_{1,2}$ are the Floquet exponents of the periodic system
  Eqs.(3) and (4). The condition for periodic ZBO is obtained when the
spacing $F|\mu_2-\mu_1|$ between the two ladders is a fractional
multiple of the single ladder spacing $Fa$.\\
In conclusion, an optical realization of coherent BO motion
superimposed to a periodic or aperiodic ZT sequence, leading to
ZBO, has been proposed. This coherent dynamics should be directly
visualized in a binary array with transverse refractive index
gradient as a sequence of beam splitting and recombination.\\
\\
Author E-mail address: longhi@fisi.polimi.it


\begin{thebibliography}{0}
\bibitem{note0}
For a curved waveguide with a local curvature $\ddot x_0$, in the
paraxial approximation and after a Kramers-Henneberger
transformation the curvature corresponds to a local index gradient
$F=-n_e \ddot x_0$ (see, for instance, Eq.(2) of \cite{Longhi05}).
For circulary-curved waveguides, in the paraxial approximation one
has $\ddot x_0= \pm 1/R$, where $R$ is the waveguide radius of
curvature, and hence $F= \mp n_e/R$. See also Ref.\cite{Lenz99}.
\bibitem{note1}
In Fig.1(b), $E_n(\kappa)$ is the eigenvalue of the Hamiltonian
operator $ \mathcal{H}=-(\hbar^2 /2 n_e) (\partial^2 / \partial
x^2)+V(x)$ with a potential $V(x)=n_e-n(x)$, where $\kappa$ in the
Bloch wave number, chosen in the first Brillouin zone, and $n$ is
the band index. Therefore, in the quantum context Fig.1(b)
represents the energy band diagram of $\mathcal{H}$, and the
energy $E_n(\kappa)$ thus increases with band index $n$. To
understand the meaning of $E_n(\kappa)$ in the optical context,
note that the effective propagation constant of the eigenmode (or
supermode) of the array for the $n$-th band is given by
$\beta_n(\kappa)=k [n_e-E_n(\kappa)]$, i.e. $n_e-E_n(\kappa)$ is
the effective index of the supermode. A different way to denote
the bands of the array, which was adopted in
Ref.\cite{Morandotti04}, is to plot the propagation constant
$\beta_n(\kappa)$ instead of $E_n(\kappa)=n_e-\beta_n(\kappa)/k$.
In this case $\beta_n(\kappa)$ decreases as the band index $n$ is
increased.
\end{thebibliography}
\end{document}